\documentstyle[aps,preprint]{revtex}
\begin{document}
\draft
\title{Charge -- Vortex Duality\\ in Double-Layered Josephson Junction Arrays}
\author{Ya.~M.~Blanter$^{a,b}$ and Gerd Sch\"on$^{c}$}
\address{$^a$ Institut f\"ur Theorie der Kondensierten Materie,
Universit\"at Karlsruhe, 76128 Karlsruhe, Germany\\
$^b$ Department of Theoretical Physics, Moscow Institute for Steel and
Alloys, Leninskii Pr. 4, 117936 Moscow, Russia\\
$^c$ Institut f\"ur Theoretische Festk\"orperphysik,
Universit\"at Karlsruhe, 76128 Karlsruhe, Germany}
\date{\today}
\maketitle
\tighten

\begin{abstract}
A system of two parallel Josephson junction arrays coupled by interlayer
capacitances is considered in the situation where one layer
is in the vortex-dominated and the other in the
charge-dominated regime. This system shows a symmetry (duality)
of the relevant degrees of freedom, i.e. the vortices in one layer and
the charges in the other.
In contrast to single-layer arrays both contribute to the kinetic energy.
The charges feel the magnetic field created by vortices,
and, vice versa, the vortices feel a gauge field created by charges.
For long-range interaction of the charges the system exhibits two
Berezinskii-Kosterlitz-Thouless transitions, one for vortices and
another one for charges. The interlayer capacitance suppresses both
transition temperatures. The charge-unbinding transition is
suppressed already for relatively weak coupling, while the
vortex-unbinding transition is more robust. The shift of the
transition temperature for vortices is
calculated in the quasi-classical approximation for arbitrary
relations between the capacitances (both weak and strong coupling).
\end{abstract}

\vspace{10pt}

Two-dimensional (2D) Josephson junction arrays have attracted much attention
because of the experimental and technological progress and the rich
underlying physics (see Ref. \cite{Mooji} for review). Classical 2D
Josephson junction arrays, where the Josephson coupling energy $E_J$
between the superconducting islands dominates, is a standard example of
a system exhibiting the Berezinskii-Kosterlitz-Thouless (BKT)
transition -- the unbinding of vortex-antivortex pairs at a certain
temperature $T_J$ \cite{BKT,Kosterlitz}. The transition separates
a superconducting phase at $T < T_J \propto E_J$, where vortices are
bound, from a resistive phase. It was realized later (see e.g. Refs.
\cite{Abeles,Doniach,LozAk,Jose1}) that charging effects,
associated with the capacitances of the islands to the ground $C_0$
and of the junctions
$C$, lead to quantum fluctuations of the phase and suppress the BKT
transition temperature. Beyond a critical value of the charging energy
$E_C = \min\{e^2/2C,e^2/2C_0\}$
the transition temperature vanishes, and the superconducting phase
ceases to exist. The next step
\cite{Fisher,Mooji2,Fazio,FGS} was to understand that in the
``extreme'' quantum limit $E_C \gg E_J$, where the quantum
fluctuations of the
phases are very strong, the vortices are ill-defined objects.
In this regime the charges on the islands become the relevant variables.
If, furthermore, $C_0 \ll C$, the interaction between the charges
is (nearly) the same as that of the vortices in the quasi-classical
array. In particular, the charges can be considered as a 2D Coulomb gas
\cite{Minn}, and they undergo a BKT transition at temperature $T_C \propto
E_C$. The phase below the transition is insulating.  A finite value
of the Josephson coupling between the islands suppresses this transition.
As another example we mention the influence of dissipation (e.g. Ohmic
dissipation) on the phase transition in the array, which was
first noted in Ref. \cite{Chakravarty}.
We are not going to review these theoretical results,
however it is necessary to stress that the theory
of 2D Josephson junction arrays is far from being settled.

Below we describe another, more complicated system - two parallel 2D
Josephson junction
arrays with capacitive coupling between them (no Josephson
coupling\footnote{Multi-layered systems {\em with} Josephson coupling
between layers have been discussed in the literature (see
e.g. \cite{Shenoy,LozKult}). The analog of the BKT transition is in this
case the disruption of vortex rings. In the limit of weak Josephson
couplings this system is reduced to the 2D XY-model, while in the
opposite case of strong coupling it is essentially the 3D XY-model.
This situation, however, is absolutely different from
the one we describe below.}). Probably the most interesting situation
arises when one array is in the quasi-classical (vortex) regime while
another one is in the quantum (charge) regime. Then the vortices in one
layer and the charges in the other one are well-defined {\em dynamical}
variables. Another important feature of the present system is that the
strength of interaction between charges and vortices is controlled by
the interlayer coupling $C_x$ and
consequently this interaction may be weak or strong, whereas in usual
2D Josephson junction arrays the strength of charge-vortex interaction is
of the same order as either the charge-charge or the vortex-vortex
interaction. We also show that the physical realization of this
interaction is rather different from that in one array. Hence, at
least for weak interlayer coupling, one should expect two BKT
transitions, the first for charges in one layer, and the second for
vortices in the other one. In this article we provide the theoretical
description of the coupled system and
calculate the  shifts of the transition temperatures
due to the interlayer interaction.

We consider two parallel Josephson junction arrays, i.e. (square)
lattices of
superconducting islands connected by Josephson links. As usual, we
suppose that the magnitude of the order parameter in the islands
is constant while its phase fluctuates from island to island.
The partition function of the system may be  expressed conveniently in
terms of these phases $\phi_{i\mu}$ (the indices $i$ label the islands
in each array and   $\mu = 1,2$ refers to the number of array)
\begin{equation} \label{partphi}
Z = \prod_{i} \int_{0}^{2\pi} d\phi_{i1}^{(0)} d\phi_{i2}^{(0)}
\sum_{\{m_{i1}\},\{m_{i2}\}} \int D \phi_{i1} (\tau) D\phi_{i2}(\tau)
\exp(-S\{\phi\})  \; .
\end{equation}
Here the path integration over phases is carried out with the
boundary conditions
$$\phi_{i\mu} (0) = \phi_{i\mu}^{(0)}; \ \ \ \phi_{i\mu} (\beta) =
\phi_{i\mu}^{(0)} + 2\pi m_{i\mu},$$
where $\beta$ is the inverse temperature, and $\hbar = 1$.
The Euclidean effective action
$S\{\phi\}$ is
\begin{eqnarray} \label{Sphi}
& & S\{\phi\} = \int_{0}^{\beta} d\tau \left\{ \frac{C_{01}}{8e^2}\sum_i
(\dot{\phi}_{i1})^2 + \frac{C_{1}}{8e^2}\sum_{\langle ij \rangle}
(\dot{\phi}_{i1} - \dot{\phi}_{j1})^2 + \frac{C_{02}}{8e^2}\sum_i
(\dot{\phi}_{i2})^2 + \frac{C_{2}}{8e^2}\sum_{\langle ij \rangle}
(\dot{\phi}_{i2} - \dot{\phi}_{j2})^2 + \right. \nonumber \\
& & \left. + \frac{C_x}{8e^2}
\sum_i (\dot{\phi}_{i1} - \dot{\phi}_{i2})^2 +
E_{J1}\sum_{\langle ij \rangle} [ 1 - \cos (\phi_{i1} - \phi_{j1})] +
E_{J2}\sum_{\langle ij \rangle} [ 1 - \cos (\phi_{i2} - \phi_{j2})]
\right\} \; .
\end{eqnarray}
Here $C_{0\mu}$ are the capacitances of the islands in the array $\mu$
relative to the ground, $C_{\mu}$ are the capacitances of the junctions in the
array $\mu$, and $C_x$ are the interlayer capacitances. Furthermore,
$E_{J\mu}$ are the Josephson coupling constants in the layers.
Here and below we use the symbol $\sum_{\langle ij \rangle}$
to denote the summation over nearest neighbors only, and
each pair is counted once; the symbol $\sum_{ij}$ stands for the
summation for all values $i$ and $j$ (in particular, each pair except
$\langle ii \rangle$ is counted twice).

{}From now on we choose parameters such
that the array 1 is in the charge (quantum)
regime while the array 2 is in the quasi-classical (vortex) regime. In
terms of the phase variables this means that in the array 1 the phases
on each grain are strongly fluctuating in time, while in array 2 they are
nearly time-independent. This regime is described by the conditions
$$E_{J1} \ll e^2/\tilde C_1,\ \ \ E_{J2} \gg e^2/\tilde C_2,$$
with $\tilde C_{\mu} = \max \{C_{0\mu}, C_{\mu}, C_x \}$. Below we
first calculate the shift of the BKT transition
temperature for vortices in the array 2. This does not require the
introduction of charges and vortices and may be done in the phase
representation. Then, we turn to the BKT transition for charges in the
array 1. For this purpose we move from a description in terms of
phases to one in terms of
 charges and vortices, and use the duality of the resulting
action to investigate the transition. At the
same time, we will show that charges and vortices in this system can
be considered as two-dimensional dynamical particles with masses. The
charge-charge
and vortex-vortex interaction are essentially those of 2D Coulomb
particles, while the charge-vortex interaction is more peculiar.

\subsection*{BKT transition for vortices}

The shift of the BKT transition temperature for vortices in the
array 2 due to the coupling to the array 1 can be calculated easily
if we set the small parameter $E_{J1}$ to zero\footnote{This means,
in particular, that
the results obtained below are valid also in the case when the array 1
is in the normal state. Because of the $e$-periodicity in this case
the boundary
conditions in the array 1 should read $\phi_{i1}(\beta) -
\phi_{i1}(0) = 4\pi m_{i1}$. As we will show, this does not
change the final result.}. Then the action for the
phases $\phi_{i1}$ becomes Gaussian and the latter may be integrated
out. After that
the shift of the BKT temperature for vortices may be obtained by means
of the quasi-classical expansion \cite{Jose1,Jose2}.

The first step requires a comment. The path integration over the
phases of the islands $\phi_{i1}(\tau)$ in the array 1 is, as usual,
performed by a linear shift of variables in order to eliminate terms
linear in $\phi_{i1}$. However, the new (shifted) variables do not
generally satisfy
the boundary conditions, and consequently the integration is not
possible. If, nevertheless, the array 2 is in the quasi-classical
regime, the contributions of all non-zero winding numbers $m_{i2}$ to
the partition function are exponentially small in comparison with the
contribution of $m_{i2}= 0$ (see, e.g. \cite{Jose2}). If we neglect
these small
contributions, the phases $\phi_{i2}$ become periodic, and the
boundary conditions in the array 1 are met automatically. After
integration over the phases $\phi_{i1}$ we find
\begin{equation} \label{part2}
Z = \prod_{i} \int_{0}^{2\pi} d\phi_{i2}^{(0)} \int D\phi_{i2}(\tau)
\exp(-\tilde
S\{\phi_{i2}\})
\end{equation}
with the effective action
\begin{equation} \label{Sphi1}
\tilde S\{\phi\} = \int_{0}^{\beta} d\tau \left\{ \frac{1}{8e^2} \sum_{ij}
\dot{\phi}_{i2} [\Lambda_2]_{ij} \dot{\phi}_{j2} +
E_{J2} \sum_{\langle ij \rangle} [ 1 - \cos (\phi_{i2} - \phi_{j2})]
\right\}.
\end{equation}
Here $[\Lambda_2]_{ij}$ is the effective capacitance matrix for the layer
2 (see also below)
\begin{equation} \label{lambda}
[\Lambda_2]_{ij} = -\frac{C_x^2}{C_1} [\hat Q_1^{-1}]_{ij} + C_2 [\hat
Q_2]_{ij}
\end{equation}
and the matrices $\hat Q_{\mu}$ ($\mu = 1,2$) have a form
\begin{eqnarray*}
[\hat Q_{\mu}]_{ij} = \left[
\begin{array}{lc}
4 + \frac{C_{0\mu} + C_x}{C_{\mu}} & i=j\\
-1 & \mbox{$i$ and $j$ are nearest neighbors}\\
0 & \mbox{otherwise}
\end{array} \right. .
\end{eqnarray*}

Since the array 2 is supposed to be in the quasi-classical regime, only
weakly time-dependent periodic paths $\phi_{i2}(\tau)$ are important. Hence we
may write the phases in the form
$\phi_{i2}(\tau) = \phi_{i2}^{(0)} + f_i(\tau)$, where
$$f_i(\tau) = \beta^{-1} \sum_{n=1}^{\infty} [f_i(\omega_n) \exp
(-i\omega_n\tau) +
f^*_i(\omega_n) \exp(i\omega_n\tau)].$$
is expressed as a sum over Matsubara frequencies $\omega_n = 2\pi n
\beta^{-1}$. Now the action may be expanded to quadratic terms in $f_i(\tau)$,
yielding
\begin{eqnarray} \label{Sphi3}
\tilde S\{\phi_{i2}^{(0)},f_i(\tau)\} & = & \beta E_{J2} \sum_{\langle ij
\rangle} [ 1 - \cos (\phi_{i2}^{(0)} - \phi_{j2}^{(0)})] + \nonumber \\
& + & \int_0^{\beta} d\tau \left\{ \frac{1}{8e^2} \sum_{ij}
\dot{f}_{i} [\Lambda_2]_{ij} \dot{f}_{j} +
\frac{E_{J2}}{2} \sum_{\langle ij \rangle} [f_i(\tau) - f_j(\tau)]^2
\cos (\phi_{i2}^{(0)} - \phi_{j2}^{(0)}) \right\}.
\end{eqnarray}
Note that the first term is the classical action of 2D Coulomb
gas \cite{Kosterlitz}. Finally, one performs the cumulant expansion
\cite{Jose1,Jose2} in
the last term in brackets in Eq. (\ref{Sphi3}). As a result the action
has exactly the same form as the classical one, but with the
renormalized temperature
\begin{equation} \label{temp1}
\beta \to \beta^{-1} - \frac{1}{2} \int_0^{\beta} d\tau \left\langle
\left( f_i(\tau) - f_j(\tau) \right)^2 \right\rangle .
\end{equation}
Here the angular brackets denote the averaging with the effective
action
$$S_{eff} = (8e^2)^{-1} \int_0^{\beta} d\tau \sum_{ij} \dot{f}_i(\tau)
[\Lambda_2]_{ij}
\dot{f}_j(\tau).$$
{}From this we obtain the shift of the transition temperature due to the
interlayer coupling
\begin{equation} \label{fin1}
T_J = T_{J0} - e^2A/3,
\end{equation}
where
$$A = \mbox{Re} ([\Lambda_2^{-1}]_{ii} - [\Lambda_2^{-1}]_{\langle ij
\rangle}),$$
and the second term is the matrix element taken for neighboring
islands. (We have assumed that the matrix $\Lambda_2$ depends
on the distance between the islands only). Here $T_{J0}$ is the
transition temperature for a classical 2D Josephson junction array
\cite{Kosterlitz} (to be of order $E_{J2}$). Eq. (\ref{fin1}) is the
result we were aiming at, however in order to obtain some analytical
expressions we evaluate the quantity $A$ in some
approximations. In Fourier representation the matrices
$Q_{\mu} (\mbox{\bf k})$ have the form
$$Q_{\mu}(\mbox{\bf k}) = (ka)^2 + (C_{\mu})^{-1}(C_x +
C_{0\mu}), \ \ \ ka \leq 1,$$
with $a$ being the lattice parameter. Consequently the matrix
$\Lambda_2^{-1}$ is
\begin{equation} \label{lambda1}
\Lambda_2^{-1} (k) = \frac{C_1(ka)^2 + C_x +
C_{01}}{(C_1(ka)^2 + C_x + C_{01})(C_2(ka)^2 + C_x + C_{02}) - C_x^2}.
\end{equation}
If we replace the first Brillouin zone by a circle cut-off at $k <
a^{-1}$, the integration over the angular variable
can be performed easily, and we obtain
$$A = \frac{a^2}{2\pi} \int_{0}^{1/a} k dk
\Lambda_2^{-1}(k) [1 - J_0(ka)].$$
In the range of integration the Bessel function $J_0$ can be
approximated by its expansion
$$J_0(x) \approx 1 - x^2/4.$$
Finally, in the case $C_{0\mu} \ll C_x$ (this situation is the most
interesting) we obtain
\begin{equation} \label{fin2}
T_{J0} - T_J = \frac{e^2}{48\pi C_{eff}}, \ \ \ \frac{1}{C_{eff}} =
\frac{1}{C_2} - \frac{C_x}{C_2^2} \ln \frac{C_xC_1 + C_xC_2}{C_xC_1 +
C_xC_2 + C_1C_2}.
\end{equation}

It is seen that the effect of layer 1 is merely the renormalization
of the effective capacitance. As a result, the BKT transition for
vortices in the layer 2 is suppressed. We should emphasize that
the result (\ref{fin2}) is valid for arbitrary capacitances $C_x$,
$C_1$ and $C_2$. The only restriction is the validity of the
quasi-classical approximation. The shift of the
transition temperature should be small, or, in other words,
$e^2/C_{eff} \ll E_{J2}$. In particular, for $C_x \ll C_2$ (weak
coupling between the layers) one obtains $C_{eff} \approx C_2$
irrespectively of $C_1$ --- vortices in layer 2 do not feel the presence
of the layer 1. In the case $C_1 = C_2 \ll C_x$ the effective
capacitance is $C_{eff} = 2C_2$, while for $C_1 = C_x \gg C_2$ one
has $C_{eff} = 2C_x/3$. It is seen that in the latter case the
temperature begins to feel the presence of the first layer, however
the absolute value of the shift becomes now small.

\subsection*{Charge -- vortex duality and BKT transition for charges}

Before we turn to the description of the BKT transition for charges,
it is necessary to stress the following. As shown by the effective
action (\ref{Sphi}), the interlayer capacitance $C_x$ not only couples
the layers, but also renormalizes the capacitances $C_{01}$ and
$C_{02}$ of the islands to the ground. Hence
the logarithmic interaction between the charges in each layer has a
finite range for any non-zero $C_x$ due to the screening, and the BKT
transition, is,
strictly speaking, absent. One should realize, however, that in the
situation $C_{01} \ll C_x \ll C_1$ the screening length $\xi_1 \sim
a(C_1/C_x)^{1/2}$ can be very large. Below we assume that
these inequalities are satisfied and the range of interaction $\xi_1$ is
large enough to make it meaningful to
speak about the charge-unbinding transition. (Note that like any phase
transition in a finite system this transition is smeared; in other
words, the resistance grows exponentially, and, strictly
speaking, for finite $\xi$ it is impossible to distinguish between
insulating and
resistive phases). On the other hand, already for relatively weak
coupling $C_x \sim C_{cr} \ll C_1$ this
description becomes meaningless and the insulating phase is
absent. As we show below, in the small range $C_x \ll C_{cr}$ the transition
temperature for charges in layer 1 does not feel the presence of the
layer 2 and hence is essentially the charge-unbinding temperature for
one Josephson junction array \cite{Fazio}. Nevertheless, the
charge-vortex description required to obtain this result gives rise to
an interesting physical model to be described below.

Now we move from the phase description (\ref{partphi}),(\ref{Sphi})
to a charge-vortex description. First we introduce the large capacitance
matrix
\begin{eqnarray} \label{matrix1}
\hat C  = \left( \begin{array}{cc}
\hat C_1 & -\hat C_x \\
-\hat C_x & \hat C_2 \\
\end{array} \right).
\end{eqnarray}
Here $\hat C_{\mu}$ is the capacitance matrix in the array $\mu$ while
$\hat C_x = C_x \delta_{ij}$. The inverse matrix in the Fourier
representation reads as
\begin{eqnarray} \label{matrix2}
& & C^{-1} (\mbox{\boldmath k}) = \frac{1}{(C_1(ka)^2 + C_x +
C_{01})(C_2(ka)^2 + C_x + C_{02}) - C_x^2} \times \\
& & \times \left(
\begin{array}{cc}
C_2 (ka)^2 + C_x +C_{02} & C_x \\
C_x & C_1 (ka)^2 + C_x +C_{01}
\end{array} \right) \equiv \left( \begin{array}{cc}
\hat \Lambda_1^{-1} (k) & \hat \Lambda_x^{-1} (k) \\
\hat \Lambda_x^{-1} (k) & \hat \Lambda_2^{-1} (k) \\
\end{array} \right) \equiv [\hat C^{-1}]^{\mu\nu}(k). \nonumber
\end{eqnarray}
Indices $\mu, \nu = 1, 2$ again label the array. The matrix $\hat
C^{-1}$ describes the interaction of charges. We have also
introduced for later convenience the matrices $\Lambda_{\mu}^{-1}$,
describing the interaction of charges within layer $\mu$, as well as
$\Lambda_x^{-1}$ referring to charges in different layers
(cf. Eq. (\ref{lambda1}). Then the effective action (\ref{Sphi}) can
be rewritten in terms of integer charges $q_{i\mu}$ of each island
and phases $\phi_{i\mu}$
\begin{eqnarray} \label{Sqphi}
S\{q,\phi\} & = & \int_{0}^{\beta} d\tau \left\{ 2e^2 \sum_{ij}
\sum_{\mu,\nu} q_{i\mu}(\tau) [\hat C^{-1}]^{\mu\nu}_{ij}
q_{j\nu}(\tau) + \sum_i \left[ q_{i1}(\tau) \dot{\phi}_{i1}(\tau) +
q_{i2}(\tau) \dot{\phi}_{i2}(\tau) \right] + \right. \nonumber \\
& & \left.
 + E_{J1}\sum_{\langle ij \rangle} [ 1 - \cos (\phi_{i1} - \phi_{j1})] +
E_{J2}\sum_{\langle ij \rangle} [ 1 - \cos (\phi_{i2} - \phi_{j2})]
\right\}.
\end{eqnarray}
Now it is possible to introduce vortex degrees of freedom by means of
the Villain transformation \cite{Villain} (see also \cite{JKKN}). It is
important that
this procedure deals only with the phase variables and does not affect
the charge interaction (the first term in Eq. (\ref{Sqphi})).
The phase terms (the second and the third one in
Eq. (\ref{Sqphi})), however, have exactly the same form in the problem
of two arrays as for a single Josephson junction layer. The procedure
for a single-layer array array is discussed in details in
Refs. \cite{Fazio,FGS}, the generalization to the double-layer system
is straightforward. The partition function becomes
\begin{equation} \label{z2}
Z = \prod_{i} \sum_{\{q_{i1},q_{i2}\}} \sum_{\{v_{i1},v_{i2}\}} \exp
(-S\{q,v\}),
\end{equation}
where the effective action for integer charges $q_{i\mu}$ and vorticities
$v_{i\mu}$ is
\begin{eqnarray} \label{Sqv1}
& & S\{q,v\} = \int_{0}^{\beta} d\tau \left\{ 2e^2 \sum_{ij}
\sum_{\mu,\nu} q_{i\mu}(\tau) [\hat C^{-1}]^{\mu\nu}_{ij}
q_{j\nu}(\tau) + \frac{1}{4\pi E_{J1} F(\epsilon_1 E_{J1})} \sum_{ij}
\dot{q}_{i1} (\tau) G^{(1)}_{ij} \dot{q}_{j1} (\tau) + \right. \nonumber \\
& & \left. + \frac{1}{4\pi E_{J2}}
\sum_{ij} \dot{q}_{i2} (\tau) G^{(2)}_{ij} \dot{q}_{j2} (\tau) + \pi E_{J1}
F(\epsilon_1 E_{J1}) \sum_{ij} v_{i1} G^{(1)}_{ij} v_{j1} + \pi E_{J2}
\sum_{ij} v_{i2} G^{(2)}_{ij} v_{j2} + \right. \\
& & \left. + i\sum_{ij} \dot{q}_{i1} (\tau)
\Theta_{ij} v_{j1} (\tau) + i\sum_{ij} \dot{q}_{i2} (\tau)
\Theta_{ij} v_{j2} (\tau) \right\} \nonumber .
\end{eqnarray}
Here we introduced the discrete time variable; the time lattice spacing
in the array $\mu$ is of order $\epsilon_{\mu} \sim
(8E_{J\mu}E_{C\mu})^{-1/2}$, $E_{C\mu} \equiv e^2/2C_{\mu}$.
The time integration and derivatives are
continuous notations for a summation over time lattice and for a
discrete derivative
$$\dot{f} (\tau) = \epsilon_{\mu}^{-1} [f(\tau + \epsilon_{\mu}) -
f(\tau)],$$
respectively. The function
$$F(x) = \frac{1}{2x \ln({J_0(x)/J_1(x)})} \to \frac{1}{2x \ln(4/x)}, \ \
\ x \ll 1,$$
is introduced to ``correct'' the Villain transformation for small
$E_J$ \cite{Villain}. As we see, its entire effect is to renormalize
(to increase) the Josephson coupling in the layer 1; the renormalized coupling
$\tilde E_{J1}$ reads as
\begin{equation} \label{EJren}
\tilde E_{J1} \sim \left(8E_{J1}E_{C1}\right)^{1/2}
\left(\ln(E_{C1}/E_{J1})\right)^{-1}.
\end{equation}
Note that for $E_{J1} \ll E_{C1}$ (charge regime) one obtains $\tilde
E_{J1} \ll E_{C1}$.

The kernel
$$\Theta_{ij} = \arctan(\frac{y_i - y_j}{x_i - x_j})$$
describes the phase configuration at site i around a vortex at site
j. Finally, the kernel $G^{(\mu)}_{ij}$ is the lattice Green's function,
i.e. the Fourier transform of $k^{-2}$. At large distances between the
sites $i$ and $j$ it depends only on the distance $r$ between the sites
and has a form (see e.g. \cite{Minn})
\begin{equation} \label{xi}
G^{(\mu)}_{ij} = \ln(\xi_{\mu}/r), \ \ \ a \ll r \ll \xi_{\mu}, \ \ \ \
\xi_{\mu} = a(C_{\mu}/C_x)^{1/2}.
\end{equation}
Later on, we assume that the linear size of each
array is much less that the range of interaction $\xi_{\mu}$. This
means, in particular, that we assume $C_x \ll C_2$.

The action (\ref{Sqv1}) depends on the charges and vorticities in both
layers. However, in our situation, when the layers 1 and 2 are in the
charge and vortex regimes, respectively, the vortices in the layer 1
and the charges in the layer 2 may be
integrated out \cite{Fazio}. To do this we suppose the latter
variables to be continuous (strongly fluctuating), and neglect the
kinetic term for charges in the layer 2 ($\dot{q}_2G^{(2)}
\dot{q}_2$). Then after performing the Gaussian integration we obtain
the effective action for charges $q_{i1}$ in the layer 1 and
vorticities $v_{i2}$ in the layer 2 (to be referred below as $q_i$ and
$v_i$)
\begin{eqnarray} \label{Sqv2}
& & S\{q,v\} = \int_{0}^{\beta} d\tau \left\{ \frac{2E_{C1}}{\pi}
\sum_{ij} q_i(\tau) G^{(1)}_{ij} q_j(\tau) + \frac{1}{4\pi \tilde
E_{J1}} \sum_{ij} \dot{q}_i (\tau) G^{(1)}_{ij} \dot{q}_j (\tau)
+ \pi E_{J2} \sum_{ij} v_i G^{(2)}_{ij} v_j + \right. \nonumber \\
& & \left. +
\frac{\pi}{8E_{C2}}
\sum_{ij} \dot{v}_i(\tau) \left[ G^{(2)}_{ij} - \frac{C_x^2}{4\pi^2
C_1C_2} \sum_{kl} \Theta_{ik} G^{(1)}_{kl} \Theta_{lj} \right]
\dot{v}_j(\tau) + \frac{iC_x}{2\pi C_1} \sum_{ijk} \dot{v}_i(\tau)
\Theta_{ik} G^{(1)}_{kj} q_j(\tau) \right\}.
\end{eqnarray}
To derive Eq. (\ref{Sqv2}) we have taken into account the explicit
expression for the large capacitance matrix (\ref{matrix2}).

The action (\ref{Sqv2}) is the central result of this section. It
looks rather similar to the effective charge-vortex action in one
Josephson junction, but the most important difference is that while in
one layer either charges or vortices are well-defined,
Eq. (\ref{Sqv2}) describes the system of {\em well-defined} dynamic
variables on each site --- charges in the layer 1 and vortices in the
layer 2. We postpone the discussion of physics in this system until
the next section, however it is clear that the action shows a duality
between charges and vortices. The second term in
the square brackets is small if $C_x \ll C_1,C_2$. Both kinetic
terms for charges and vortices violate the duality due to the
numerical coefficients. However close enough
to the transitions these terms produce only small renormalization of
the transition temperature, and are not important.
Another interesting feature of this action is that the last term,
describing the
interaction between charges and vortices, is also small, while in a
single-layer array the interaction is always of the same order of
magnitude as another terms.

It is obvious that for long-range interaction of the charges in the
layer 1 they also exhibit the BKT
transition, and under the conditions where the action (\ref{Sqv2}) was
obtained the transition temperature does not feel the presence of the
layer 2:
$$T_{C0} - T_C = \frac{\tilde E_{J1}}{24\pi}.$$

\subsection*{Charge and vortex motion}

To understand the physics described by the action (\ref{Sqv2}) it is
instructive to map this model onto the 2D Coulomb gas. For this purpose
we move from the space-time lattice to the continuous medium and
introduce the coordinates of the vortex centers and charges
\begin{eqnarray} \label{decompose}
& & q_i(\tau) \to \sum_{m} q_m \delta(\mbox{\bf r} - \mbox{\bf
r}_m(\tau)) \nonumber \\
& & v_i(\tau) \to \sum_{n} v_n \delta(\mbox{\bf r} - \mbox{\bf
R}_n(\tau)).
\end{eqnarray}
Here $q_m = \pm 1$ and $v_n = \pm 1$ represent charges and
vortices respectively; $\mbox{\bf r}_m(\tau)$ and $\mbox{\bf
R}_n(\tau)$ are the corresponding coordinates of the charges and of
the vortex centers. Now the partition function reads
\begin{equation} \label{part3}
Z = \sum_{M=0}^{\infty} \sum_{N=0}^{\infty} \int D\mbox{\bf r}_1(\tau)
\dots D\mbox{\bf r}_{2M}(\tau) D\mbox{\bf R}_1(\tau) \dots D\mbox{\bf
R}_{2N}(\tau) \exp(-S\{\mbox{\bf r},\mbox{\bf R} \}),
\end{equation}
and we are going to deal with the effective action $S\{\mbox{\bf
r},\mbox{\bf R} \}$, describing the behavior of the system of $2M$
charges (of which $M$ are positive, $q=1$, and the other $M$ are negative,
$q=-1$), and of $N$ positive ($v=1$) and $N$ negative ($v=-1$)
vortices.

The first and third terms of the action (\ref{Sqv2}) can be easily
transformed by means of decomposition (\ref{decompose}). The first one
produces the potential energy of charge interaction,
\begin{equation} \label{s1}
S_{int}^{(q)} = \frac{2E_{C1}}{\pi} \int_0^{\beta} d\tau
\sum_{m,n=1}^{2M} q_m q_n G^{(1)} (\mbox{\bf r}_m(\tau) - \mbox{\bf
r}_n(\tau)).
\end{equation}
In principle, the summation includes the terms with $m = n$; these,
however, may be excluded from this sum, giving rise to the chemical
potential for charges. The third term in Eq. (\ref{Sqv2}) yields the
interaction of vortices
\begin{equation} \label{s2}
S_{int}^{(v)} = \pi E_{J2} \int_0^{\beta} d\tau \sum_{m,n=1}^{2N} v_m
v_n G^{(1)} (\mbox{\bf R}_m(\tau) - \mbox{\bf R}_n(\tau)).
\end{equation}
Here again the term with $m = n$ gives rise to the chemical potential
for vortices. The terms (\ref{s1}) and (\ref{s2}) are essentially the
action for (classical) Coulomb gases of charges and vortices,
respectively \cite{Kosterlitz}.

If we neglect the small correction proportional to the $C_x^2/C_1C_2$
in the fourth term in Eq.(\ref{Sqv2}) then the second and fourth terms
can be transformed to the kinetic energy of charges and vortices
respectively \cite{Fazio}. The second term gives
\begin{equation} \label{s31}
S_{kin}^{(q)} = \frac{1}{2\pi \tilde E_{J1}} \int_0^{\beta} d\tau
\sum_{m,n=1}^{2M} q_m q_n \dot{r}_m^{\gamma} M_{\gamma\delta}
(\mbox{\bf r}_m - \mbox{\bf r}_n) \dot{r}_n^{\delta}.
\end{equation}
We have introduced the mass tensor \cite{Eckern}
\begin{equation} \label{mass}
M_{\gamma\delta} (\mbox{\bf r}) = -\nabla_{\gamma} \nabla_{\delta}
G^{(\mu)}(\mbox{\bf r}).
\end{equation}
It decreases proportional to $r^{-2}$ for $r \gg a$, and consequently may
be approximated by a local function
$$M_{\gamma\delta} (\mbox{\bf r}) = M\delta_{\gamma\delta}
\delta(\mbox{\bf r}), \ \ \ M = \frac{\pi}{a^2}.$$
Then the kinetic term for charges takes a simple form
\begin{equation} \label{s3}
S_{kin}^{(q)} = \frac{1}{2 a^2 \tilde E_{J1}} \int_0^{\beta} d\tau
\sum_{m=1}^{2M} \dot{r}_m^2 (\tau).
\end{equation}

Similarly, the fourth term in Eq.(\ref{Sqv2}) produces the
kinetic term for vortices
\begin{equation} \label{s4}
S_{kin}^{(v)} = \frac{\pi^2}{8 a^2 E_{C2}} \int_0^{\beta} d\tau
\sum_{m=1}^{2N} \dot{R}_m^2 (\tau).
\end{equation}

Finally, the last term in Eq.(\ref{Sqv2})) is responsible for the
interaction between charges and vortices. The corresponding term in
S\{\mbox{\bf r},\mbox{\bf R}\} is
\begin{equation} \label{s51}
S_{qv} = \frac{iC_x}{2\pi C_1 a^2} \int_{0}^{\beta} d\tau \sum_{mn}
v_m q_n \int d\mbox{\bf r}' \nabla_{\mbox{\bf R}_m} \Theta  (\mbox{\bf
R}_m - \mbox{\bf
r}') G^{(1)} (\mbox{\bf r}' - \mbox{\bf r}_n) \dot{\mbox{\bf R}}_m(\tau).
\end{equation}
The integral over $\mbox{\bf r}$ can be calculated explicitly,
yielding
\begin{equation} \label{s5}
S_{qv} = - \int_{0}^{\beta} d\tau \sum_{m} i v_m \dot{\mbox{\bf
R}}_m(\tau) \mbox{\bf A}_m (\mbox{\bf R}_m)
\end{equation}
with
$$\mbox{\bf A}_m (\mbox{\bf R}_m) = \sum_n q_n \mbox{\bf a} (\mbox{\bf
R}_m (\tau) - \mbox{\bf r}_n (\tau)),$$
$$\mbox{\bf a} (\mbox{\bf r}) = - \frac{1}{4a^2} \frac{C_x}{C_1} (1 +
2\ln(\xi_1/r)) [\hat z \times \mbox{\bf r}].$$
The resulting action is
\begin{equation} \label{fin}
S\{\mbox{\bf r},\mbox{\bf R}\} = S_{int}^{(q)} + S_{int}^{(v)} +
S_{kin}^{(q)} + S_{kin}^{(v)} + S_{qv}.
\end{equation}

The action (\ref{fin}) is essentially that of two 2D Coulomb systems. The
charges and the vortices can be viewed as particles with masses
\begin{equation} \label{mass1}
M_q =  \frac{1}{a^2\tilde E_{J1}} \ \ \mbox{and} \; \;
M_v =  \frac{\pi^2}{4a^2 E_{C2}} \ \,
\end{equation}
respectively. Charges interact via the effective capacitance, vortices
via the usual
logarithmic interaction with strength $E_{J2}$. Furthermore, the vortices
produce the vector potential $\mbox{\bf a}$ for the
charges\footnote{This seeming
asymmetry is rather artificial. In Eq.(\ref{Sqv2}) one can rewrite
after a partial integration the
charge-vortex interaction term in order to obtain the vector potential
for vortices, created by charges, as
well. However, this vector potential contains always the small factor
$C_x/C_1$.}; the magnetic field associated with
this vector potential is
\begin{equation} \label{field}
B = \pm \frac{1}{ea^2} \frac{C_x}{C_1} \ln \frac{\xi_1}{r},\ \ \ a \ll
r \ll \xi_1.
\end{equation}
Its sign depends of the signs of the corresponding vortex and
charge. Apart from its quite peculiar functional form, another
important feature of this field is the small factor $C_x/C_1$.

\subsection*{Summary}

We have investigated the system of two
2D Josephson junction arrays coupled by capacitances $C_x$, in the
situation when the arrays 1 and 2 are in the charge and vortex regime,
respectively. In the case of
weak coupling $C_x \ll C_1,C_2$ the system shows an (approximate)
duality between dynamical charges in one layer and dynamical vortices
in the other one. In contrast to a single layer array,
both variables are well-defined. The system is equivalent to two
2D Coulomb gases of massful particles. The charges feel
the magnetic field created by vortices, and, vice versa, the vortices
feel the gauge field created by charges. In this respect the system
resembles the composite fermion model of the fractional quantum Hall
effect, however the magnetic field is now small and has another
functional form, so one may expect different physics.
In this regime
the system shows two BKT transitions, one for charges and another for
vortices, and the coupling between the layers suppresses both
transitions. Although one could expect the suppression of one
transition and the enhancement of another one, the suppression of both
transitions is rather natural, since the capacitance $C_x$ also
renormalizes the capacitances of the islands to the
ground. The BKT transition for charges vanishes even for very small
values of $C_x$, however, the BKT
transition for vortices survives under condition $e^2/C_{eff} \ll
E_{J2}$ irrespective of the relations between the capacitances
$C_1$, $C_2$ and $C_x$. The shift of this temperature due to the
capacitance effects is calculated within the phase
representation for both cases of weak and strong coupling.
The effect of the layer 1 is to renormalize the
capacitance matrix in the layer 2. For weak coupling $C_x \ll C_2$
(irrespective of $C_1$) the vortices do not feel the presence of
another layer, and the temperature remains the same as for one
layer. However, for $C_x \gg C_2$ different situations are possible.

In summary, we would like to emphasize that the system of two coupled
Josephson junction arrays may exhibit quite rich and interesting
physics. We have investigated some limiting cases, however, the further
rich behavior of this system  can be expected in other cases.
In particular, the magnetic field created by vortices seems
to be rather unusual and interesting. We hope that experimental
studies of this system will be performed in the near future.

\subsection*{Acknowledgments}

The authors are grateful to R.~Fazio, K.-H.~Wagenblast, and
A.~D.~Zaikin for useful discussions. One of us (G.~S.) acknowledges
the hospitality of the Helsinki University of Technology, where the
part of this work was done. The project was supported
by the Alexander von Humboldt Foundation (Y.~M.~B.) and by the DFG
within the research program of the Sonderforschungbereich 195.

\end{document}